\documentstyle[a4wide,epsfig,11pt,titlepage]{article}

\newcounter{nref}
\newcommand{\bbib}{%
  \renewcommand{\refname}{\large\bf References}%
  \begin{thebibliography}{99}%
  \setcounter{enumiv}{\thenref}}
\newcommand{\ebib}{%
  \end{thebibliography}%
  \setcounter{nref}{\arabic{enumiv}}} 
\newcommand{\head}[3]{%
  \setcounter{nref}{0}%
  \thispagestyle{empty}%
  \section*{\LARGE\bf #1}%
  \stepcounter{section}%
  \addcontentsline{toc}{section}{#1}%
  \large\itshape%
  #2\\\vspace{0.1pt}\\%
  #3%
  \normalsize\upshape%
  \bigskip}

\begin{document}


\head{Model atmospheres for type~Ia supernovae:\\
  Basic steps towards realistic synthetic spectra}
{D.N.\ Sauer$^{1,2}$, A.W.A\ Pauldrach$^{2}$}
{$^1$ Max-Planck-Institut f\"ur Astrophysik, Garching, Germany\\
 $^2$ Institut f\"ur Astronomie und Astrophysik der Universit\"at M\"unchen,
Germany}

\subsection{Introduction}

Analyses and interpretations of the meanwhile enormous database of
excellent spectra of type~Ia supernovae obtained in key projects and
extensive searches still suffer from missing realistic numerical
simulations of the physics involved. One of the most important steps
towards this objective concerns a detailed simulation of the radiative
transfer in the outermost parts of this objects in order to calculate
synthetic spectra on the basis of models of the explosion scenario.
The priority objective thereby is to construct for the first time
consistent models which link the results of the hydrodynamics and
nucleosynthesis obtained from the explosion models (see, for example,
\cite{dsauer.reinecke01}, \cite{dsauer.reinecke99}, these proceedings)
with the calculations of light curves (see, for instance,
\cite{dsauer.blini00},\cite{dsauer.soro00}, these
proceedings) and synthetic spectra of SN~Ia. After a phase of testing
this new tool will provide a method for SN~Ia diagnostics, whereby
physical constraints on the basic parameters, the abundances, and the
hydrodynamic structures of the atmospheric part of the objects can be
obtained via a detailed comparison of observed and synthetic spectra.
This will also allow to determine the astrophysically important
information about the distances of SN~Ia.

First results in this direction are presented in Section 1.3. In the preceding
sections we will first with regard to our objective briefly summarize the most
important observed properties of type~Ia supernovae, and then discuss the
current status of our treatment of the hydrodynamic expanding atmospheres of
SN~Ia.

\subsection{Observed properties of type~Ia supernovae}

The classification of supernovae is based on purely empiric
observational properties. In this scheme, type~Ia supernovae are
defined by the absence of hydrogen and helium spectral lines and a
strong absorption feature of Si\,{\sc ii} lines near 6100\,\AA\ in
epochs around maximum light. In late phases the SN~Ia spectra are
dominated by emission of forbidden lines of mainly iron and cobalt.

Prominent features in the early-time spectra of type~Ia supernovae --
this is the phase we want to analyze at present -- are characteristic
absorption lines resulting from low ionized intermediate-mass elements
(such as Si\,{\sc ii}, O\,{\sc i}, Ca\,{\sc ii}, Mg\,{\sc ii}). The
formation of these lines results in a pseudo-continuum that is set up
by the overlap of thousands of these lines.  The absorption lines show
characteristic line shapes due to Doppler-broadening resulting from
large velocity gradients. In the UV-part of the spectrum, the flux is
strongly depleted (compared to a blackbody-fit in the optical range)
by line blocking of heavy element lines (especially Fe, Co, Mg).
Towards the red and IR, the flux is also low due to the absence of
significant line influence in these regions. Towards later phases
($\sim 2$ weeks after max.), heavier element lines become more
prominent as the photosphere recedes deeper into the ejecta. At this
time emission features also start to increasingly dominate the
spectral characteristics. In the nebular phase ($\sim 1\,$month) the
spectrum approaches a typical nebular appearance dominated by lines of
heavy element forbidden transitions.

The spectra as well as the light curves of type~Ia supernovae show a
high degree of homogeneity; however, some intrinsic differences are
also observed. The mechanisms leading to these differences have to be
understood in order to draw conclusions based on the homogeneity, such
as the distance measurements for cosmological applications.  (For
detailed reviews of the optical properties of SN~Ia we refer to
\cite{dsauer.leibu00},\cite{dsauer.filip97}.)

\subsection{Model atmospheres for type~Ia supernovae}

In early phases, which are considered here, a supernova of type~Ia is
described by a photosphere in the deeper layers (i.e., the sphere
where the optical depth $\tau=1$) and a superimposed extended
atmosphere that expands at high velocities -- up to 30000\,km/s.

Due to the high radiation energy density and the dominant role of
scattering processes the problem we are dealing with is strongly
non-local, leading to typical conditions that make consideration of
NLTE (non local thermodynamic equilibrium) effects necessary.
Additionally, processes of line blocking and blanketing significantly
influence the radiative transfer within the ejecta and thus the
observed spectral properties (see \cite{dsauer.pauldrach96}). This
behavior results in the strong depletion of the flux in the UV part
and the blue region of the spectrum due to the influence of thousands
of lines of the heavy elements. As the absorbed radiation is partly
reemitted towards lower layers (``{\em backwarming\/}'') the
temperature structure is also affected, and in turn the ionization
equilibrium and thus the radiation field itself.

Certain properties of supernovae make the full problem of NLTE and
radiative transfer calculations even more complicated than in related
objects, such as hot stars: The element abundances of type~Ia
supernovae are entirely dominated by heavy species that have fairly
complicated ionic energy level structures with hundred-thousands of
transitions. In addition, the high velocities within the ejecta
broaden these lines by Doppler-shift and thus cause a strong overlap
of sometimes thousands of lines especially in the UV wavelength
regions. In contrast to objects with similar physical conditions (for
example, hot stars or, in late phases, planetary nebulae), SNe are not
illuminated by a central source of radiation, but are heated by the
diffuse source of the $\gamma$-rays emitted by the radioactive decay
of $^{56}$Ni and $^{56}$Co. Thus, the diffuse field is of primary
importance and can not be approximated accurately in a simple way.
Another problem concerns the relatively flat density distribution:
Even for early phases where the assumption of an underlying
photosphere seems to be justified, the radius of this photosphere
(defined as the radius where the optical depth $\tau _\nu=1$) varies
strongly with wavelength.

The approach used in this project employs a detailed atmospheric model
code which is based on the concept of {\em homogeneous, stationary,
  and spherically symmetric radiation-driven winds}, where the
expansion of the atmosphere is due to scattering and absorption of
Doppler-shifted metal lines. This code provides a detailed and
consistent solution of the radiative transfer including a proper
treatment of NLTE as well as blocking and blanketing effects in order
to calculate realistic synthetic spectra (see
\cite{dsauer.pauldrach01}, \cite{dsauer.pauldrach96}). The computation
of the NLTE level populations is carried out using detailed atomic
models, and including all important contributions to the rate
equations. Thomson scattering, bound-free and free-free opacities,
line absorption and emission processes as well as dielectronic
recombination is included in the radiative transfer. The computation
of the NLTE model is based on several iteration cycles of step-wise
improving accuracy.  In a first step, an opacity sampling method is
employed, which results in a approximate solution that is, however,
already close to the final solution. An exact treatment of the
radiative transfer equation in the observer's frame is performed in a
second step. In a final step the occupation numbers, opacities and
emissivities from the NLTE model are utilized for calculating a
synthetic spectrum via a formal solution of the transfer equation. For
a detailed description of the numerical methods we refer to
\cite{dsauer.pauldrach01}.

\subsection{Results}

We regard the results obtained so far as not yet suitable for
diagnostic purposes, since some physics relevant for SN~modelling have
been only very roughly included up to now (e.g., the heating by the
diffuse source of the $\gamma$-rays emitted by the radioactive decay
of $^{56}$Ni and $^{56}$Co).  Thus, our present calculations should be
considered as preliminary. All models obtained so far are based on the
hydrodynamic model W7 by Nomoto et al.~\cite{dsauer.nomoto84}. Since
radial variation of the chemical composition has not yet been fully
implemented in our models described here, the values used for the
composition have been averaged over mass shells. Additionally we
assume that the energy deposition arises completely from the optically
thick part of the atmosphere. Moreover, the expansion of the ejecta is
assumed to be homologous (i.e. $r\propto v$).

\begin{figure}[ht]
   \centerline{%
     \epsfxsize=0.5\textwidth\epsffile{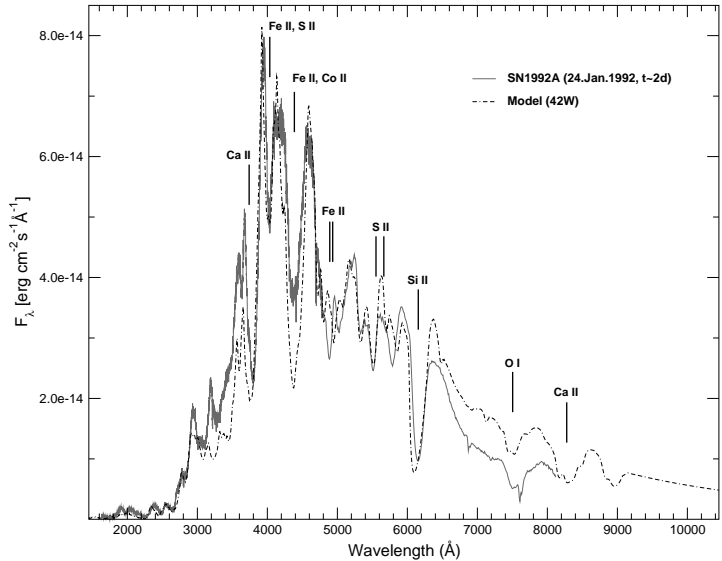}}
   \caption{Model calculation compared to observed spectrum of SN1992A}
   \label{dsauer.fig1}
\end{figure}

Fig.~\ref{dsauer.fig1} compares a calculated model spectrum to the
observed ``standard'' type~Ia supernova 1992A spectrum.  As is shown,
the synthetic spectrum reproduces the observed spectrum in the UV and
blue wavelength ranges.  Thus, our method already produces
quantitatively reliable results indicating that the basic physics is
obviously treated properly. However, towards the redder part of the
spectrum a systematic offset to the observation appears.  We regard
our treatment of the inner boundary condition (diffusion
approximation) and the missing energy deposition above the photosphere
as a possible reason for this behavior. At the inner boundary we apply
the diffusion approximation at all frequencies. This approximation is
valid for conditions where the mean free path of the photons is very
small (i.e. $\tau \gg 1$), which, however, is not guaranteed at longer
wavelengths, due to the absence of strong lines in this wavelength
range and the relatively flat density distribution that does not
provide the required strong increase of the continuum opacities
towards our innermost radius grid point. As a consequence, the flux
emitted from the inner boundary is too high. In addition, the opacity
in this wavelength range is dominated by pure electron (Thomson)
scattering that does not couple the radiation field to the thermal
pool. In these regions the radiation observed is actually generated in
much deeper layers of the ejecta and scattered outwards.

\subsection{Near future prospects}

Regarding a solid basis for spectral diagnostics of type~Ia supernovae
the next steps are obvious. Depth-dependent abundances and a proper
treatment of the $\gamma$~energy deposition above the photosphere have
to be included. The latter point will also free us from the limitation
of currently being able to treat only the photospheric, early-phase
epochs around maximum.  With these enhancements it seems feasible to
obtain synthetic spectra based on hydrodynamic explosion models like
those of the hydrodynamics group at MPA, and to compare them directly
to observed SN~spectra.

\subsection*{Acknowledgements}

We wish to thank our collaborators at MPA -- the Hydro-Group -- and
USM for helpful discussions. In particular we thank T.~Hoffmann at USM
for providing lots of useful tools and active support for this ongoing
project. The first author thanks especially his primary supervisor,
W.~Hillebrandt, for numerous helpful comments, permanent motivation,
and financial support.



\end{document}